\documentclass[12pt]{iopart}
\usepackage{amssymb}
\usepackage{graphics}
\begin{document}
\title{Spectral Line Removal in the LIGO Data Analysis System (LDAS)}
\author{Antony C Searle, Susan M Scott and David E McClelland}
\address{Department of Physics, Faculty of Science, The Australian National University, Canberra ACT 0200, AUSTRALIA}
\eads{\mailto{antony.searle@anu.edu.au}, \mailto{susan.scott@anu.edu.au}, \mailto{david.mcclelland@anu.edu.au}}
\begin{abstract}
High power in narrow frequency bands, \emph{spectral lines}, are a feature of an interferometric gravitational wave detector's output.  Some lines are coherent between interferometers, in particular, the 2~km and 4~km LIGO Hanford instruments.  This is of concern to data analysis techniques, such as the stochastic background search, that use correlations between instruments to detect gravitational radiation.  Several techniques of `line removal' have been proposed.  Where a line is attributable to a measurable environmental disturbance, a simple linear model may be fitted to predict, and subsequently subtract away, that line.  This technique has been implemented (as the command \texttt{oelslr}) in the LIGO Data Analysis System (LDAS).  We demonstrate its application to LIGO S1 data.
\end{abstract}
\submitto{\CQG}
\maketitle
\section{Introduction}

\begin{figure}
\resizebox{\textwidth}{!}{\includegraphics{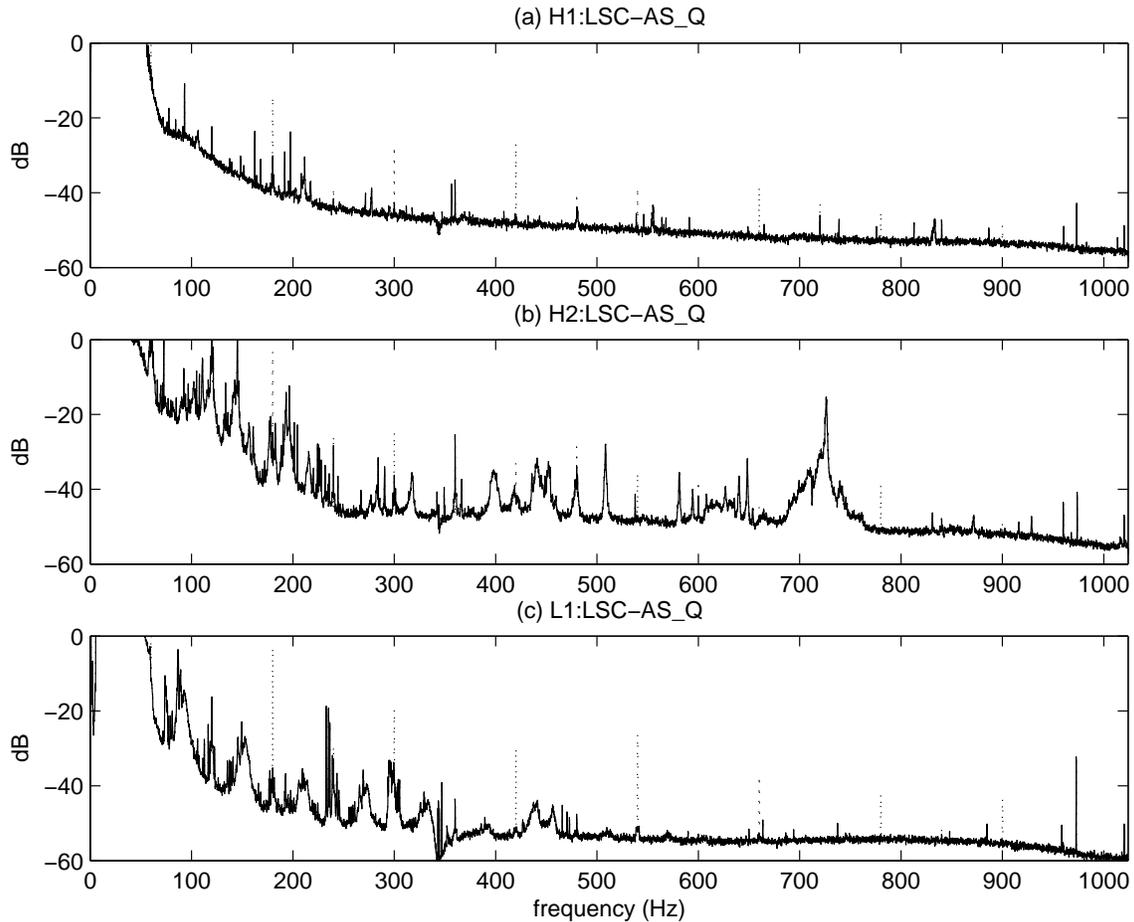}}
\caption{LIGO Hanford Observatory (a) 4~km interferometer (H1) and (b) 2~km interferometer (H2) and (c) LIGO Livingston Observatory 4~km interferometer (L1) output power spectra (uncalibrated), before (dotted line) and after (solid line) application of the line removal technique described in \S\ref{sec:implementation}, for GPS times 714975000--714975600.}
\label{fig:psd}
\end{figure}

\begin{figure}
\resizebox{\textwidth}{!}{\includegraphics{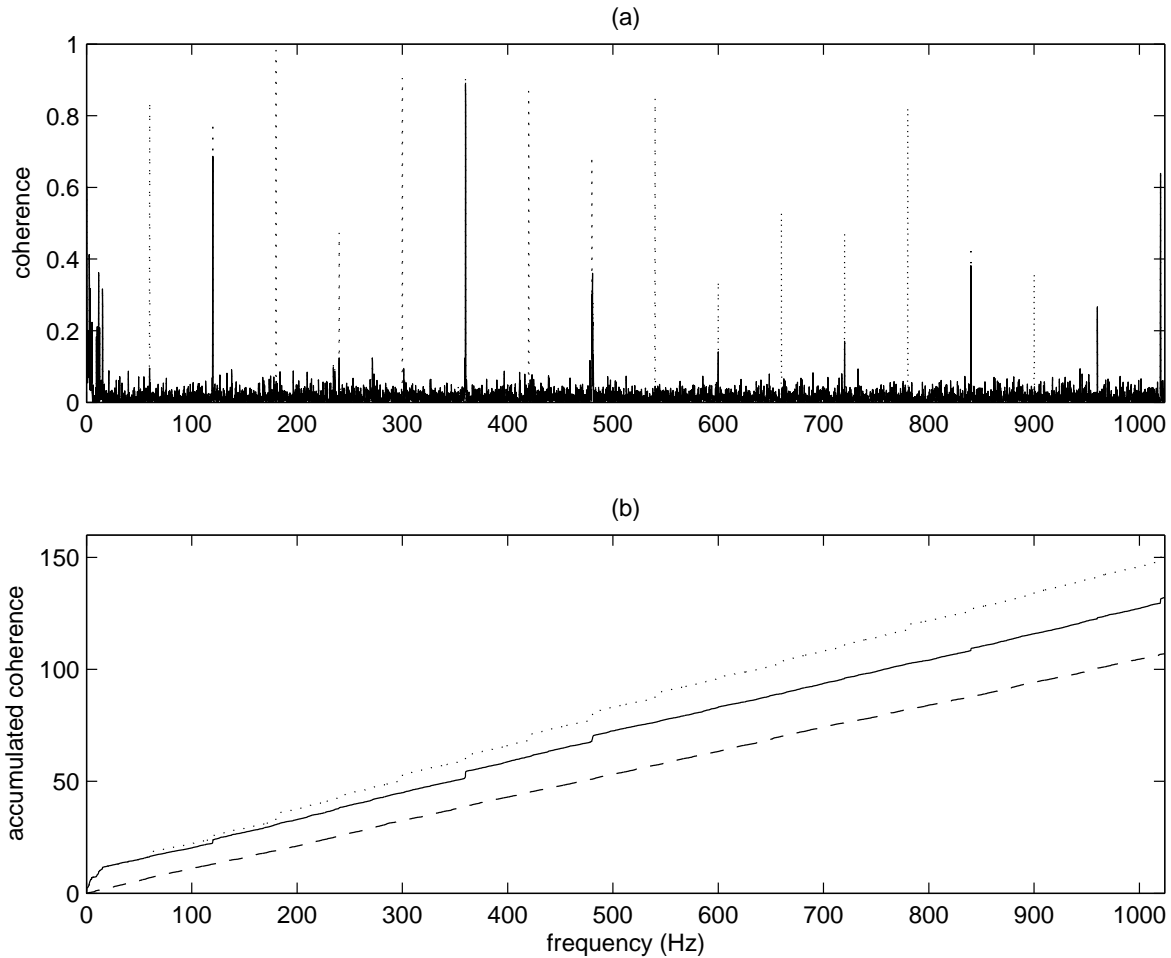}}
\caption{(a) Coherence of H1:LSC-AS\_Q and H2:LSC-AS\_Q before (dotted line) and after (solid line) application of the line removal technique described in \S\ref{sec:implementation}, for GPS times 714975000--714975600. (b) Accumulated coherence of H1:LSC-AS\_Q and H2:LSC-AS\_Q before (dotted line) and after (solid line) application of the line removal technique, with the accumulated coherence of H1:LSC-AS\_Q and L1:LSC-AS\_Q (dashed line) provided for reference, for GPS times 714975000--714975600.}
\label{fig:cohereh1}
\end{figure}

The broadband sensitivity of the nascent instruments of the Laser Interferometer Gravitational-wave Observatory (LIGO) \cite{ligo} is dictated by seismic, thermal and shot noise.  There are also, however, a number of narrow-band noise sources---spectral lines.  Resonances of the mirror suspension wires (\emph{violin modes}) are one example of these.  The extreme sensitivity of the LIGO instruments makes them particularly susceptible to contamination from the environment.  Seismic noise, dictating the lower limit on the frequency of detectable astrophysical sources is perhaps the most dramatic example.  Similarly, the 60 Hz frequency of the United States (US) electrical supply, and its harmonics, form a comb of strong spectral lines in each LIGO instrument, across the entire observation band (Figure~\ref{fig:psd}).  These lines are the largest single factor in the coherence between the 2~km and 4~km LIGO Hanford interferometers (Figure~\ref{fig:cohereh1}).

Spectral lines render data non-Gaussian, increase the required dynamic range of measurements, and obscure possible gravitational wave signals at their own and nearby frequencies.  Some environmental noise sources can introduce spurious correlations between interferometers.  A number of techniques have been proposed to \emph{remove} such spectral lines from datasets using a variety of strategies.  Many concentrate on producing a model of a line, so that it may be subtracted out. Sintes and Schutz \cite{sintes} have implemented a technique to exploit common information between line harmonics to build such a line model.  Allen, Hua and Ottewill \cite{allen} have used frequency-domain linear regression against channels measuring disturbances to the instrument to remove cross-talk between channels.

LIGO's searches for astrophysical sources of gravitational radiation are predominantly performed by the LIGO Data Analysis System (LDAS) \cite{ldas}.  LDAS runs on networks of computers at a number of sites, including the LIGO Observatories, and partner institutions, including The Australian National University (ANU), where the Australian Consortium for Interferometric Gravitational Astronomy (ACIGA) maintains the ACIGA Data Analysis Cluster (ADAC).  LDAS encompasses data acquisition, processing and storage.  Architecturally, LDAS is divided into a number of inter-communicating processes that run on server machines, which collectively support parallel search codes on Beowulf clusters.  The general-purpose Data Conditioning API (datacondAPI) supports a \textsc{Matlab}-like scripting language to perform general non-parallel pre-processing \emph{actions} on incoming data before it is distributed to the cluster.  Spectral line removal is one of the available actions.

In LDAS, line removal is implemented in the datacondAPI, using the \texttt{oelslr} (Output Error Least Square Line Removal) action.  It may be added, in a straightforward manner, to the existing datacondAPI script for an astrophysical search, as another pre-processing stage in the analysis `pipeline'.  An optional line removal step has already been added to the data conditioning script for the stochastic background search.

The obvious figure of merit for line removal is the change it produces in the sensitivity of the astrophysical searches whose data it pre-processes.  At this time we do not present any such results (though in the case of the stochastic background search they are forthcoming \cite{stochastic}).
There is, as yet, no general test to determine the quality of a line removal algorithm, either in an absolute sense, or with respect to another line removal algorithm.  The reduction of correlation, coherence or non-gaussianity, or the recovery of injected signals, have variously been employed to test a technique \cite{sintes, allen}.  We demonstrate both the reduction of coherence and (via power spectra) the reduction of non-gaussian features.  (We intend to demonstrate the recovery of injected signals and employ further gaussianity testing \cite{charlton} in a later paper.)  These two effects can reasonably be expected to benefit astrophysical searches.  Conversely, line removal can be used to establish that a given astrophysical search is \emph{not} adversely affected by the presence of spectral lines (see \cite{stochastic} for a detailed discussion of this point in the context of the stochastic background search).

\begin{figure}
\resizebox{\textwidth}{!}{\includegraphics{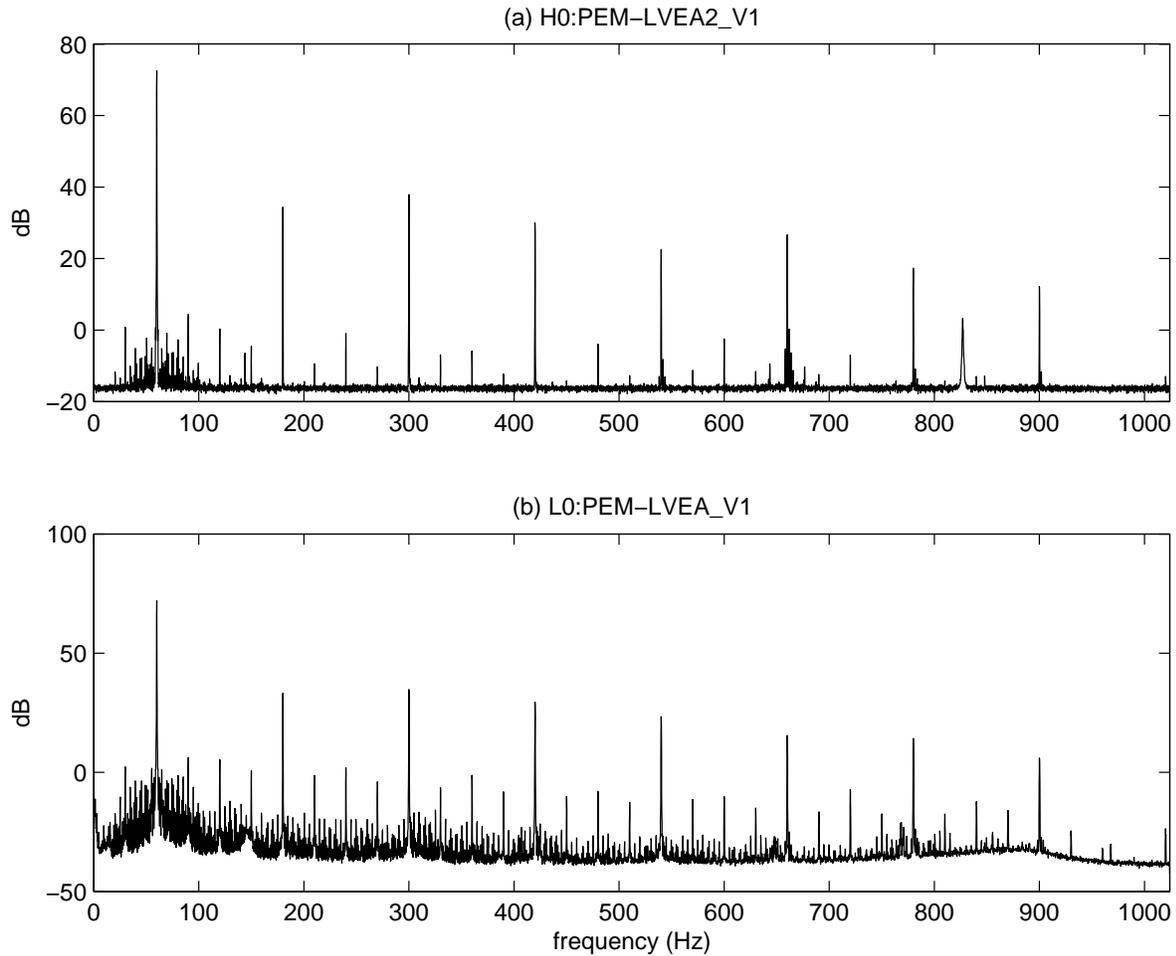}}
\caption{(a) LIGO Hanford Observatory and (b) LIGO Livingston Observatory voltage monitor channel (uncalibrated) power spectra for GPS times 714975000--714975600.}
\label{fig:psdv}
\end{figure}

Our implementation is intended as the first step in adding line removal functionality to the datacondAPI.  Similarly to Allen \etal \cite{allen}, it makes use of the additional information available in the form of a measurement of the disturbance to the instrument---the voltage monitor channels recorded by an observatory (Figure~\ref{fig:psdv}).  With a known, measured disturbance, we may use system identification theory techniques \cite{ljung} to construct a \emph{model} of the interferometer's response to the disturbance from actual data recorded over some epoch.  At subsequent times, the model can then be used to predict the interferometer's response from the measured disturbance; this prediction can then be subtracted away to remove those lines from the interferometer's output.

\section{Implementation\label{sec:implementation}}
The Finite Impulse Response (FIR) model is the simplest special case of a \emph{regression model}\footnote{Line removal in the datacondAPI was originally designed to use the Output-Error (OE) model, but the complexity of an OE model estimator has delayed its implementation.}, where the (possibly complex) input $u$ and output $y$ of a system are related by (following Ljung \cite{ljung}, where $t$ is a discrete sample number)
\begin{equation}
y(t) \approx b_1u(t-1)+b_2u(t-2)+\cdots+b_m u(t-m),\ b_1\cdots b_m\in{\mathbb C},
\end{equation}
where the model for the system is $\theta=[b_1\ldots b_{m}]^T$.  The model order $m$ controls the range of samples contributing to the model output.

If we define $\varphi(t)=[u(t-1)\ldots u(t-m)]^T$, then $\hat{y}(t|\theta)=\varphi^T(t)\theta$ is the modelled output of the system.  For given $y$ and $u$, the best model $\hat{\theta}_N$ (which minimises, over $\theta$, the sum over $N$ samples of the square of the prediction error),
\begin{equation}
\hat{\theta}_N = \arg\min_\theta\sum_{t=1}^N[y(t)-\varphi^T(t)\theta]^2,
\end{equation}
can be determined analytically (where $\overline{\varphi(t)}$ denotes the complex conjugation of $\varphi(t)$):
\begin{equation}
\hat{\theta}_N=\left[\sum_{t=1}^N\overline{\varphi(t)}\varphi^T(t)\right]^{-1}\sum_{t=1}^N y(t)\varphi(t).
\end{equation}

For an FIR model with a white noise (or \emph{error}) term, $e(t)$, such that
\begin{equation}
y(t)=b_1u(t-1)+b_2u(t-2)+\cdots+b_m u(t-m)+e(t),
\end{equation}
it is important to note that $\hat{\theta}_N$ is an unbiased estimate of $\theta$, converging as $N^{-\frac{1}{2}}$ \cite{ljung}.

In the context of gravitational wave data analysis, we assume that the detector output $y(t)$ would consist of white noise $e(t)$, except for the presence of another channel $u(t)$ that linearly and additively contaminates it with $b_1u(t-1)+b_2u(t-2)+\ldots+b_m u(t-m)$.  To remove the contamination, we estimate ${b_i}$ and subtract away the model prediction from the measured data:
\begin{equation}
y_{\mathrm{r}}(t)=y(t)-\varphi^T(t)\hat{\theta}_N,
\end{equation}
where $y_\mathrm{r}(t)$ is the system output with the line removed.

Where the contamination is localised in frequency space, as for spectral lines, we perform band selection in the time domain using pre-existing components of the datacondAPI.  If $u(t)$ is sampled with a Nyquist frequency $f_{\mathrm{Ny}}$ and a line is contained in some interval $f\pm f_\mathrm{Ny}/n$ (where $n$ is a \emph{downsampling ratio}; see below), then $u$ is first mixed down to zero frequency with multiplication by $e^{-i2\pi ft/f_\mathrm{Ny}}$.  The datacondAPI's non-trivial \texttt{resample} algorithm \cite{resample} is then used to downsample to a new time series with Nyquist frequency $f_\mathrm{Ny}/n$.  Note that \texttt{resample} does not simply return every $n$th element of a time series; it includes filtering stages to prevent the aliasing of high-frequency components into the result. Restricting the bandwidth to an integer fraction of the Nyquist frequency allows the time-domain resampling to be performed efficiently.  The downsampling ratio $n$ may typically be quite large---we use 128 in the following analysis---to identify a narrow band.  A side-effect of the resampling algorithm is that any sequences processed by the line remover must contain an integer number of $n$ samples, making it advantageous to have $n$ correspond to some common divisor of the desired sample counts---typically a power of 2.  It is on this pre-processed version of $u$ that the model is fitted.  The process is reversed to produce $\hat{y}$ by upsampling (again including a filtering stage) and up-mixing the model prediction\footnote{This series of operations has been abstracted into a reusable C++ class \cite{stroustrup} (\texttt{datacondAPI::BandSelector}) that has already been used in the implementation of the datacondAPI's Kalman filter.}.

In the datacondAPI, line removal is performed using the \texttt{oelslr} action to both estimate $\theta$ and predict $\hat{y}$.  Note that, like its underlying C++ implementation \cite{stroustrup}, the datacondAPI command language allows \emph{overloading} of function names, so that \texttt{oelslr} may perform different tasks depending on the number and type of arguments supplied to it.
\begin{eqnarray}
\theta=\mathtt{oelslr(}y\mathtt{,}u\mathtt{,}\frac{f}{f_\mathrm{Ny}}\mathtt{,}n\mathtt{,}m\mathtt{);}\\
\hat{y}=\mathtt{oelslr(}u\mathtt{,}\theta\mathtt{);}
\end{eqnarray}
The general purpose \texttt{sub} action is used to perform the final subtraction.
\begin{equation}
y_\mathrm{r}=\mathtt{sub(}y\mathtt{,}\hat{y}\mathtt{);}
\end{equation}
Time-domain causal linear filters introduce start-up transients and time-delays.  The filters are employed both in the band-selection (via \texttt{resample}) and the model implementation.  This introduces start-up transients and time-delays into the prediction.  The start-up transients invalidate the first $nm$ samples of the prediction.  The band-selection also truncates the prediction by an implementation-defined multiple of $n$ samples ($128n$ samples for the current \texttt{resample} implementation).  These issues can be simply addressed by providing $u$ for $[t_1 - \delta t, t_2 + \delta t)$ where $\hat{y}$ is required for $[t_1, t_2)$; for typical parameters, $\delta t = O(\mathrm{seconds})$.

A line removal LDAS `job' is composed as follows.  An estimation era is \texttt{slice}d from incoming channels (aliased \texttt{u} and \texttt{y}).
\begin{verbatim}
            ue = slice(u, 0, 1228800, 1);
            ye = slice(y, 0, 1228800, 1);
\end{verbatim}
where
\begin{equation}
\mathtt{slice(}y\mathtt{,}a\mathtt{,}b\mathtt{,}c\mathtt{)} = z(t) = y(a+ct),\ t=0,1,\cdots,b-1.
\end{equation}
We have `sliced off' $b=1228800$ samples, starting from $a=0$, with stride $c=1$.  For these 2048 Hz channels, this corresponds to the first 10 minutes of data for each channel.  To remove the 180 Hz line, we estimate a model \texttt{theta} from \texttt{ye} and \texttt{ue}, in a region of $(\mathtt{0.17578125}\pm\mathtt{128}^{-1})f_\mathrm{Ny}$, with model order $m=8$.
\begin{verbatim}
            theta = oelslr(ye, ue, 0.17578125, 128, 8);
\end{verbatim}
The next stage is to \texttt{slice} a prediction era from \texttt{u}.  We wish to predict $\hat{y}$ for the subsequent 10 minutes, and allowing (for simplicity) a generous $\delta t$ of 1 minute, we use a slice of \texttt{u} extending for 12 minutes from 9 minutes after the channel began taking data.
\begin{verbatim}
            up = slice(u, 1105920, 1474560, 1);
\end{verbatim}
To produce a prediction \texttt{yp} requires only the model \texttt{theta} and the predictor \texttt{up}.
\begin{verbatim}
            yp = oelslr(up, theta);
\end{verbatim}
Once the prediction has been produced, we reset it to a slice of its own middle 10 minutes (beginning 1 minute into the $\lessapprox12$ minute raw prediction), effectively discarding the start-up transients and trailing truncation.
\begin{verbatim}
            yp = slice(yp, 122880, 1228800, 1);
\end{verbatim}
We store the measured values of \texttt{y} for the corresponding times in \texttt{ym}, in preparation for subtracting the prediction from the measurement to produce the line-removed sequence \texttt{yr}.
\begin{verbatim}
            ym = slice(y, 1228800, 1228800, 1);
            yr = sub(ym, yp);
\end{verbatim}
The sequence \texttt{yr} can then be output to other LDAS APIs for further processing by astrophysical searches, or, as in \S\ref{sec:results}, written to file for inspection.

\section{Results\label{sec:results}}
Data taken during LIGO runs is formatted as \emph{frames} of named channels.  The `gravitational wave channels' of the 4 km (H1) and 2 km (H2) Hanford and 4 km (L1) Livingston observatories are, respectively, \texttt{H1:LSC-AS\_Q}, \texttt{H2:LSC-AS\_Q} and \texttt{L1:LSC-AS\_Q}. In this paper, we have used data from the S1 Science Run, specifically from a stretch of `triple-coincidence' data from GPS times 714974400--714975660, during which time all interferometers were locked; all figures are produced using the 10 minutes of data from GPS times 714975000--714975600.

The \texttt{LSC-AS\_Q} channels of all three interferometers show spectral lines at multiples of 60 Hz in their power spectra (Figure~\ref{fig:psd}), particularly for odd harmonics (60 Hz, 180 Hz, 300 Hz, \ldots).  These lines are strongly coherent between H1 and H2 (Figure \ref{fig:cohereh1}), but not between H1 and L1 or H2 and L1 (not shown).  The lines are attributed to interference from the 60 Hz alternating current mains supply, and the coherence is attributed to the fact that a common mains supply is shared between H1 and H2 at Hanford, WA, but not by L1 at Livingston, LA, as the electrical grid is not coherent between the sites on this short timescale \cite{coherence}.

\begin{figure}
\resizebox{\textwidth}{!}{\includegraphics{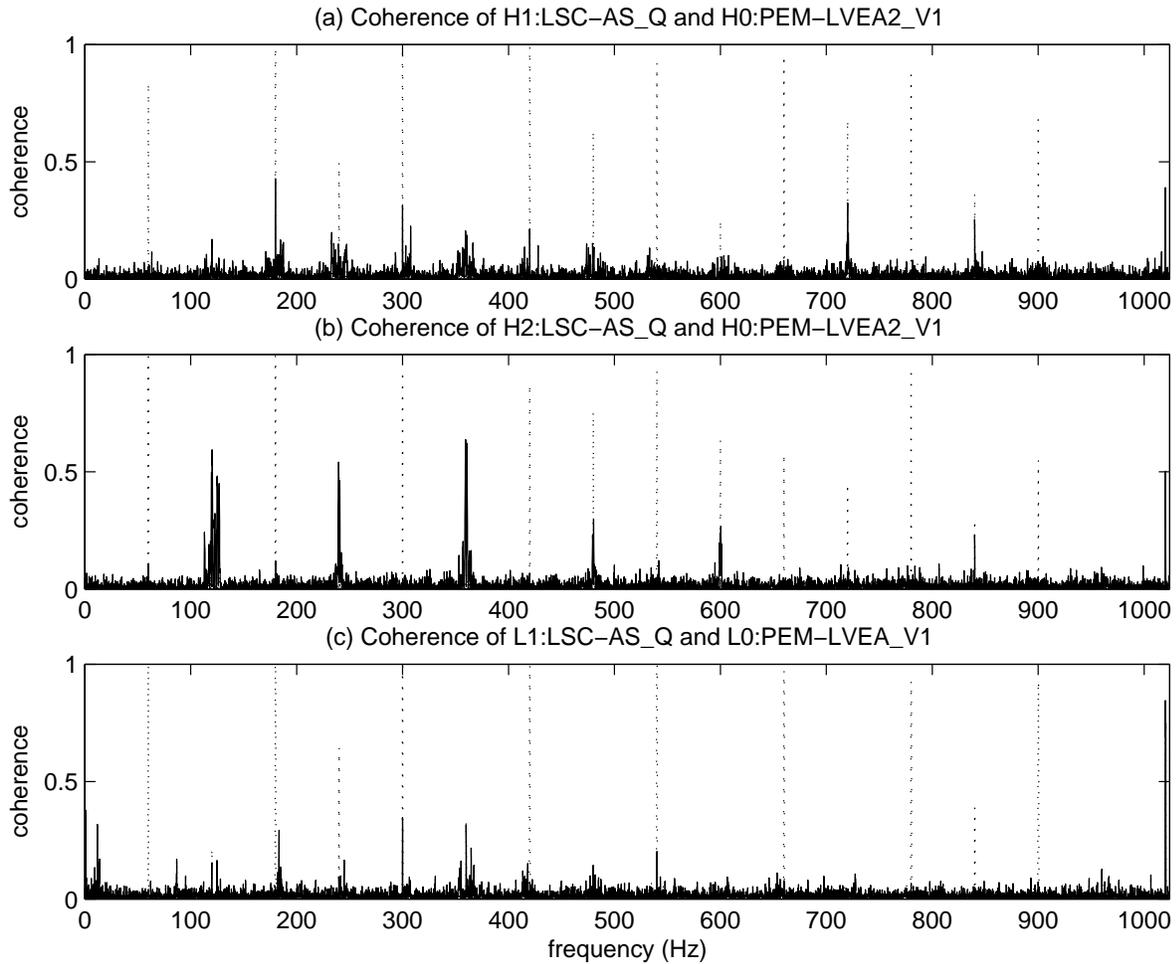}}
\caption{Coherence of (a) H1:LSC-AS\_Q, (b) H2:LSC-AS\_Q and (c) L1:LSC-AS\_Q with their respective voltage monitor channels, H0:PEM-LVEA2\_V1 and L0:PEM-LVEA\_V1, before (dotted line) and after (solid line) application of the line removal technique described in \S\ref{sec:implementation}.}
\label{fig:coherev}
\end{figure}

The power spectra of the mains voltage at both Hanford and Livingston Observatories (Figure~\ref{fig:psdv}) exhibit prominent lines at the odd harmonics of 60 Hz, and weaker lines at the even harmonics.  At each observatory, the voltage of the incoming mains supply is measured by several monitors and recorded to corresponding channels; we use \texttt{H0:PEM-LVEA2\_V1} and \texttt{L0:PEM-LVEA\_V1}.  The \texttt{LSC-AS\_Q} channels are typically strongly coherent with their local voltage monitors at odd harmonics, and weakly coherent at even harmonics (Figure~\ref{fig:coherev}). This indicates that the voltage monitor channels should be good predictors of the odd harmonics, and fair predictors of the even harmonics, given the simple linear model used by the line remover.  Despite this, our results should be treated with caution.  While it is clear that the electrical supply is responsible for the lines we observe at multiples of 60 Hz, if it is due to some non-linear effects then our description of the system is flawed.

Assuming that we may regress \texttt{*:LSC-AS\_Q} against \texttt{*0:PEM-LVEA*\_V1}, we construct an LDAS job to separately remove lines at each of the 17 multiples of 60 Hz below the 1024 Hz Nyquist frequency of the 2048 Hz \texttt{*0:PEM-LVEA*\_V1} channels (the \texttt{*:LSC-AS\_Q} channels are downsampled from 16384 Hz to 2048 Hz before this stage of the data conditioning).  The technique proceeds for each line as in \S\ref{sec:implementation}, with the exception that for each gravitational wave channel we store a single prediction sequence consisting of the sum of the predictions for each of the lines for that channel---a single unified $\hat{y}$ predicting all lines.  The parameters $m$ and $n$ used for the line removal have, anecdotally, been found to produce reasonable results; they have not been optimised for this particular task.

\begin{figure}
\resizebox{\textwidth}{!}{\includegraphics{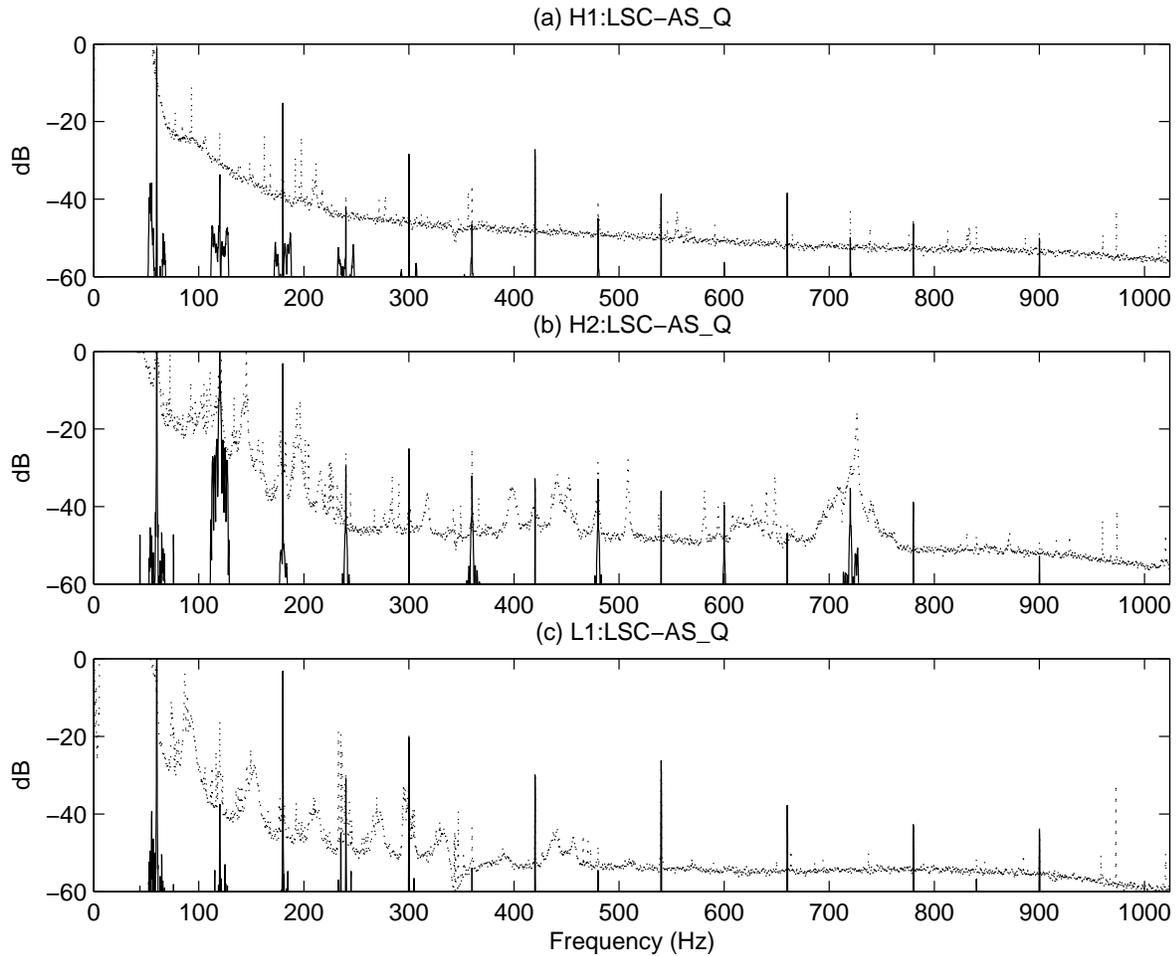}}
\caption{Power spectra of the predictions for (a) H1:LSC-AS\_Q, (b) H2:LSC-AS\_Q and (c) L1:LSC-AS\_Q (solid line).  Corresponding power spectra of the gravitational wave channels are provided for reference (dotted line).}
\label{fig:psdp}
\end{figure}

Several features are noteworthy on the power spectra of the predictions (Figure~\ref{fig:psdp}).  Firstly, the power of the prediction never exceeds the power of the \texttt{*:LSC-AS\_Q} channel.  Outside the bands selected for line removal, the power is 5--10 orders of magnitude below the \texttt{*:LSC-AS\_Q} noise floor.  Within the bands selected for line removal, the power (other than at the spectral line itself) is at least 2 orders of magnitude below the \texttt{*:LSC-AS\_Q} noise floor.  Most importantly, for the lines themselves, the power of the prediction is comparable to the power of the lines.

When the prediction is subtracted from the measured gravitational wave channel, the lines, as measured in the power spectra of Figure~\ref{fig:psd}, are affected to varying degrees.  Many are no longer visible above the noise floor; others have been reduced but are still present; some are unaffected.  The residual coherence (Figure~\ref{fig:coherev}) between the line removed channels and their predictors is similar.  For most lines, the coherence has been reduced; for many there is no residual coherence above the noise level of the estimate.

Similarly, the line coherence between the H1 and H2 interferometers (Figure~\ref{fig:cohereh1}) has been reduced or removed for almost all lines.  (The coherence between H1 or H2 and L1, not shown, is unaffected.)  This can be seen, most clearly, by considering the accumulated coherence; lines appear as steps in the accumulation.  For H1 and H2, those steps have been reduced or eliminated.  Furthermore, the accumulation demonstrates that there has been no significant broadband coherence added to the interferometers.  The net effect has been a reduction in the total accumulated coherence between the interferometers, and a significant one, when compared to the noise floor approximated by the accumulated H1--L1 coherence.

\section{Conclusion}
The datacondAPI \texttt{oelslr} line removal algorithm provides a non-intrusive way for LIGO's astrophysical searches to reduce the power and inter-instrument coherence of spectral lines attributed to interference from the mains supply.  The datacondAPI implementation provides an easy route for searches to add line removal, and provides the building blocks for expansions to the datacondAPI, some of which have already been reused in a Kalman filter implementation.

Much remains to be done.  The behaviour of the \texttt{oelslr} algorithm needs to be systematically investigated, to fully validate its implementation and to optimise its application.  Concerns about the linearity and causality of the relationship between the environmental and detector channels---and hence the validity of the model---must be addressed.  Search code integration and performance testing have already begun, and the results of one such investigation are forthcoming \cite{stochastic}.

\ack We gratefully acknowledge the assistance of the LDAS development team. This work was supported by The ANU Faculties Research Grant Scheme and an award under the Merit Allocation Scheme on the National Facility of the Australian Partnership for Advanced Computing.  A Searle was supported by an Australian Postgraduate Award.

\section*{References}
\bibliography{acsearle}
\bibliographystyle{unsrt}

\end{document}